\documentclass[12pt]{article}
\usepackage{graphicx}

\newcommand\pubnumber{CMS CR-2017/370}
\newcommand\pubdate{\today}

\def\institute{Institut f\"ur Experimentelle Teilchenphysik\\
Karlsruher Institut f\"ur Technologie, 76131 Karlsruhe, GERMANY}

\def\Title#1{\begin{center} {\Large #1 } \end{center}}
\def\Author#1{\begin{center}{ \sc #1} \end{center}}
\def\Address#1{\begin{center}{ \it #1} \end{center}}

\newcommand\pubblock{\rightline{\begin{tabular}{l} \pubnumber\\
         \pubdate  \end{tabular}}}
\newenvironment{Abstract}{\begin{quotation}  }{\end{quotation}}
\newenvironment{Presented}{\begin{quotation} \begin{center} 
             PRESENTED AT\end{center}\bigskip 
      \begin{center}\begin{large}}{\end{large}\end{center} \end{quotation}}

\def\beq{\begin{equation}}
\def\eeq#1{\label{#1}\end{equation}}
\def\eeqn{\end{equation}}

\def\beqa{\begin{eqnarray}}
\def\eeqa#1{\label{#1}\end{eqnarray}}
\def\eeqan{\end{eqnarray}}

\let\bar=\overbar

\def\Dslash{\not{\hbox{\kern-4pt $D$}}}
\def\dslash{\not{\hbox{\kern-2pt $\del$}}}

\def\msb{{\bar{\ssstyle M \kern -1pt S}}}

\newcommand{\sigmat}{\ensuremath{\sigma_{t\textrm{-ch.},\textrm{t}+\bar{\textrm{t}}}}}
\newcommand{\sigmattop}{\ensuremath{\sigma_{t\textrm{-ch.,t}}}}
\newcommand{\sigmatantitop}{\ensuremath{\sigma_{t\textrm{-ch.,}\bar{\textrm{t}}}}}
\newcommand{\xsectheotop}{ 136.02 }
\newcommand{\xsectheoantitop}{ 80.95 }
\newcommand{\xsectheo}{ 216.99 }
\renewcommand{\xsectheotop}{ 136.0 }
\renewcommand{\xsectheoantitop}{ 81.0 }
\renewcommand{\xsectheo}{ 217.0 }
\newcommand{\xsectheotopscale}{\ensuremath{^{+4.09}_{-2.92}}}
\newcommand{\xsectheoantitopscale}{\ensuremath{^{+2.53}_{-1.71}}}
\newcommand{\xsectheoscale}{\ensuremath{^{+6.62}_{-4.64}}}
\renewcommand{\xsectheotopscale}{\ensuremath{^{+4.1}_{-2.9}}}
\renewcommand{\xsectheoantitopscale}{\ensuremath{^{+2.5}_{-1.7}}}
\renewcommand{\xsectheoscale}{\ensuremath{^{+6.6}_{-4.6}}}
\newcommand{\xsectheotoppdf}{\ensuremath{\pm3.52}}
\newcommand{\xsectheoantitoppdf}{\ensuremath{\pm3.18}}
\newcommand{\xsectheopdf}{\ensuremath{\pm6.16}}
\renewcommand{\xsectheotoppdf}{\ensuremath{\pm3.5}}
\renewcommand{\xsectheoantitoppdf}{\ensuremath{\pm3.2}}
\renewcommand{\xsectheopdf}{\ensuremath{\pm6.2}}

\begin{document}
\begin{titlepage}
\pubblock

\vfill
\Title{Cross section measurement of $t$-channel single top quark production in pp collisions at $\sqrt{s}$ = 13 TeV}
\vfill
\Author{Nils Faltermann \\on behalf of the CMS Collaboration}
\Address{\institute}
\vfill
\begin{Abstract}
Single top quarks in the $t$ channel are produced though the electroweak force and are therefore an excellent way to probe the electroweak sector of the standard model with top quark physics. The recent cross section measurement of single top $t$-channel production from the CMS collaboration is presented in this poster. The analyzed data was taken during the LHC Run~II in 2015 at a center-of-mass energy of $13\,\textrm{TeV}$. A multivariate classifier is used to distinguish signal and background processes and the cross section is then extracted using a binned maximum-likelihood fit.
\end{Abstract}
\vfill
\begin{Presented}
$10^{\mathrm{th}}$ International Workshop on Top Quark Physics\\
Braga, Portugal, September 17--22, 2017
\end{Presented}
\vfill
\end{titlepage}
\def\thefootnote{\fnsymbol{footnote}}
\setcounter{footnote}{0}

\section{Introduction}

\begin{figure}[htb]
\centering
\includegraphics[height=1.5in]{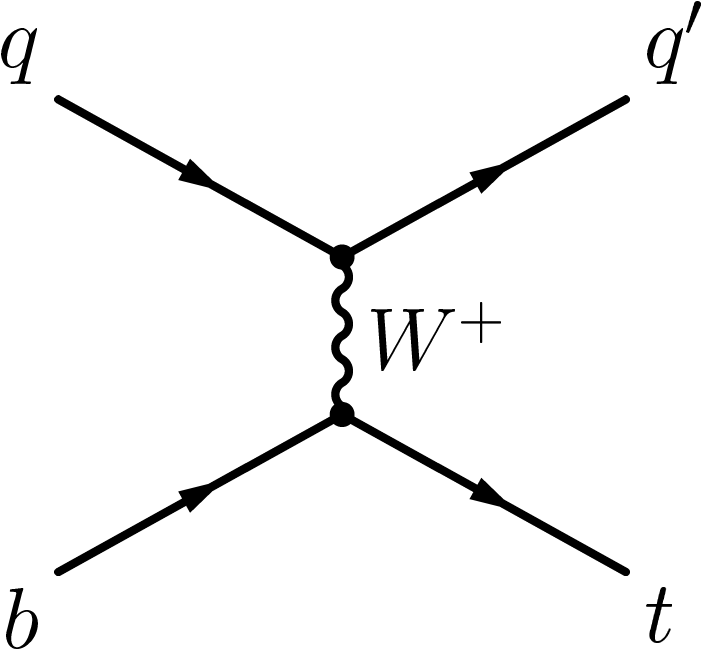}\qquad\qquad
\includegraphics[height=1.5in]{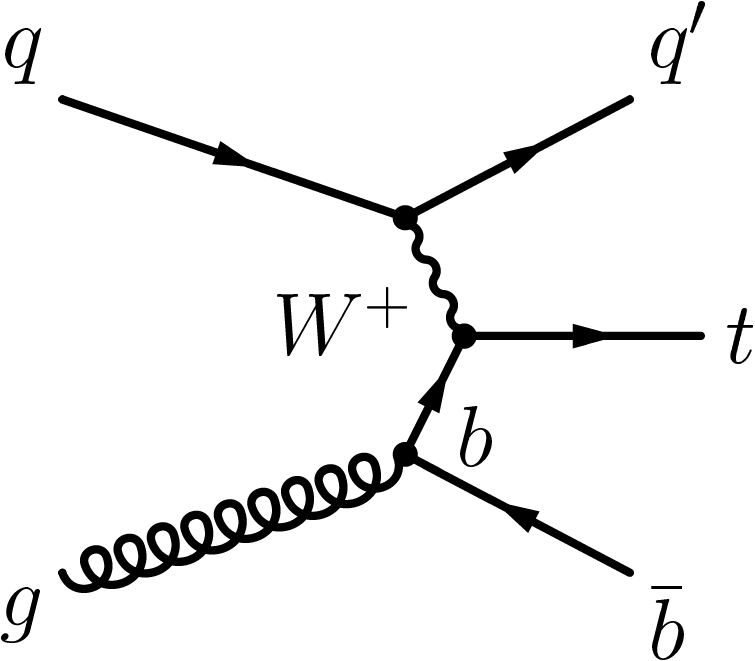}
\caption{Feynman diagrams in leading order for the $t$-channel production of single top quarks. The five-flavor scheme is shown on the left, where the bottom quark comes directly from the initial proton. In the four-flavor scheme, which is shown on the right, the bottom quark is created from an initial gluon splitting.}
\label{fig:fdia}
\end{figure}

Although top quarks are dominantly produced in pairs at the LHC, single top quark production has a significant contribution to the overall top quark production. The top quark pair production is mediated via the strong force, while for single top quark production it is the electroweak force through a $\mathrm{Vtb}$ vertex. At a center-of-mass energy of $13\,\textrm{TeV}$ about $70\%$ of all single top quarks are produced via the $t$ channel. The process can theoretically described either in the five-flavor scheme, where the bottom quark is one of the initial partons ($2 \rightarrow 2$ process), or the four-flavor scheme, where the bottom quark emerges from an additional gluon splitting. Figure~\ref{fig:fdia} shows the leading order Feynman diagrams for both flavor schemes. Other production mechanisms are the tW-associated production and the $s$-channel production, which account for the reaming $30\%$ of the single top quark production rate.
The predicted cross sections at $13\,\textrm{TeV}$ for top quark, antitop quark and inclusive $t$-channel production at next-to-leading order, calculated using {\sc HATHOR}~\cite{hathor}, are:
\begin{eqnarray*}
\sigmattop &=& \xsectheotop  \, \xsectheotopscale\,{\rm (scale)} \xsectheotoppdf \,(\textrm{PDF}{+}{\alpha_{\rm S}})\,\textrm{pb}, \\
\sigmatantitop &=&  \xsectheoantitop \, \xsectheoantitopscale\, {\rm (scale)} \xsectheoantitoppdf\,(\textrm{PDF}{+}{\alpha_{\rm S}})\,\textrm{pb}, \\
\sigmat &=&  \xsectheo \, \xsectheoscale\, {\rm (scale)} \xsectheopdf \,(\textrm{PDF}{+}{\alpha_{\rm S}})\,\textrm{pb}
\end{eqnarray*}
\section{Measurement}
\begin{figure}[htb]
\centering
\includegraphics[height=1.8in]{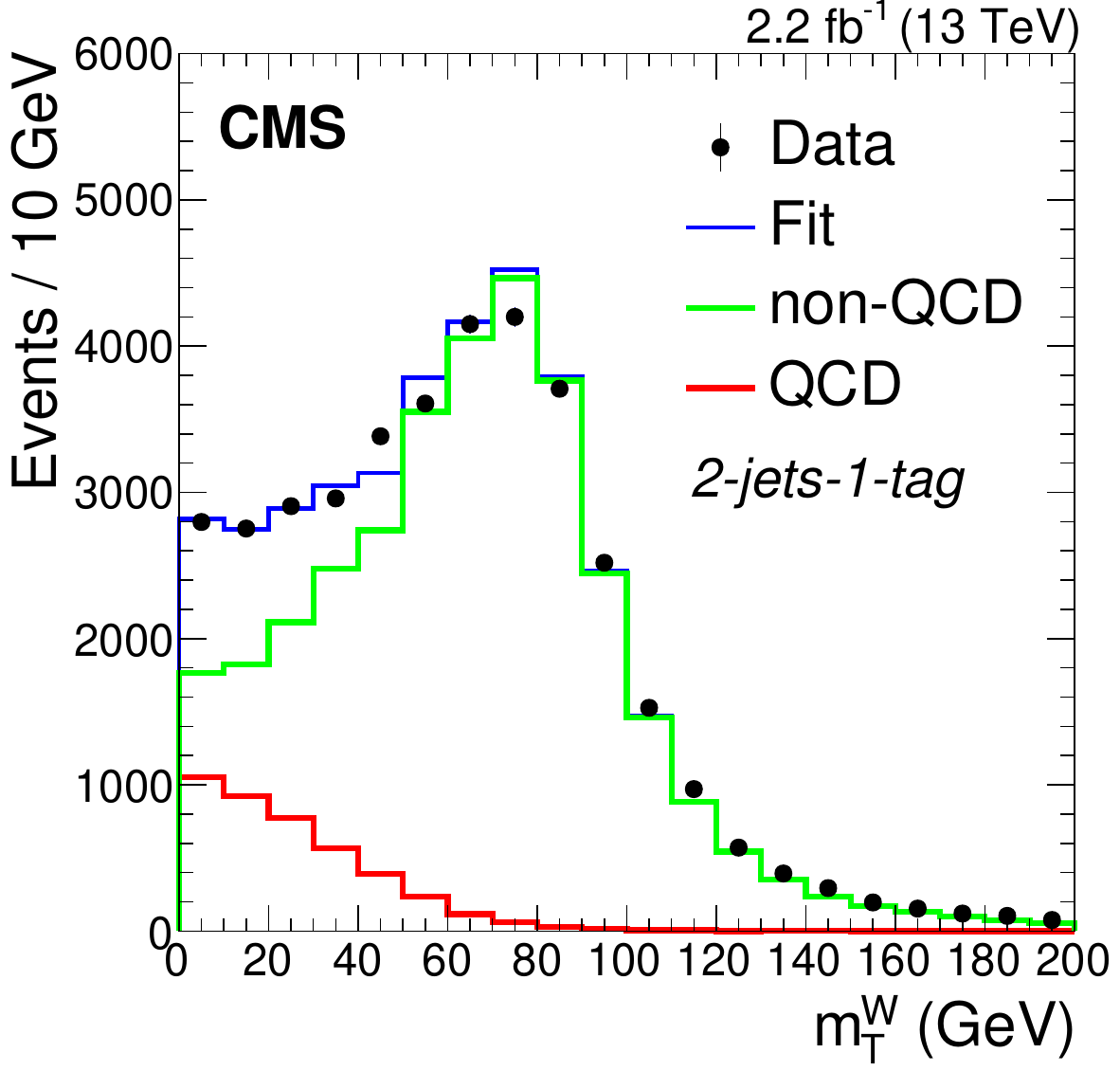}
\includegraphics[height=1.8in]{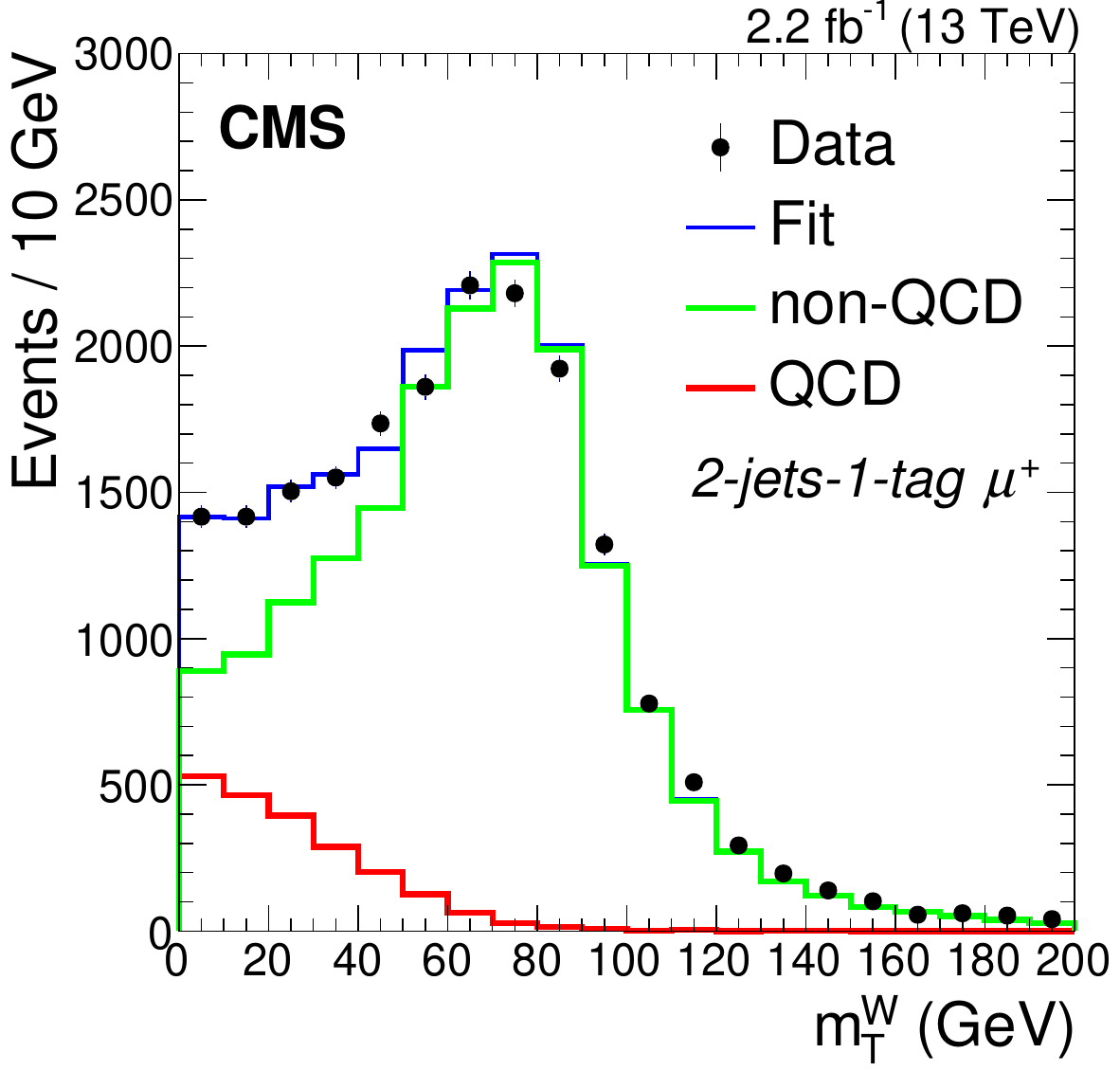}
\includegraphics[height=1.8in]{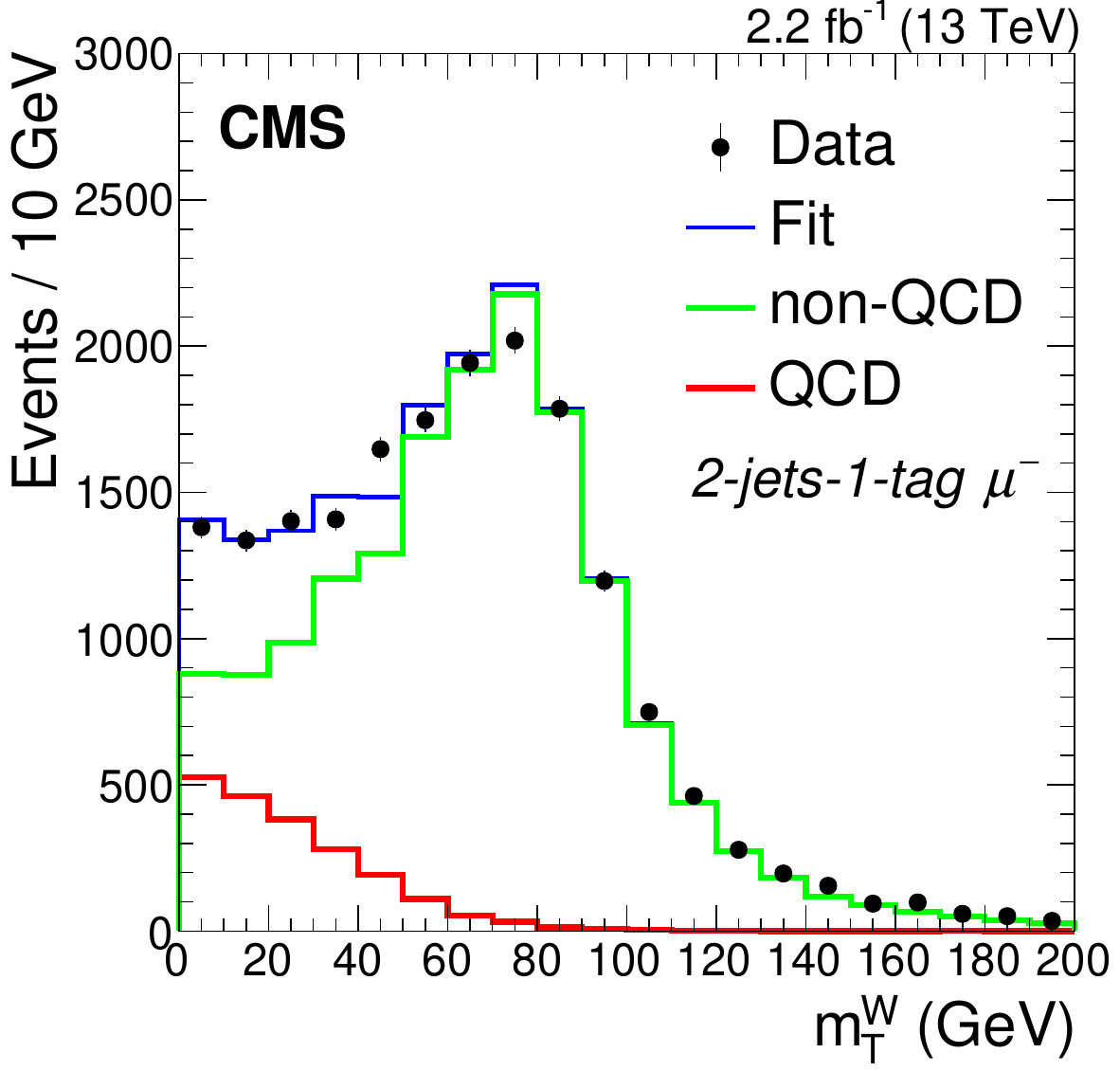}
\caption{QCD estimation in the main signal region for the inclusive selection (left), only positively charged muons (middle) and only negatively charged muons (right)~\cite{paper}.}
\label{fig:qcd}
\end{figure}

\begin{figure}[htb]
\centering
\includegraphics[height=1.8in]{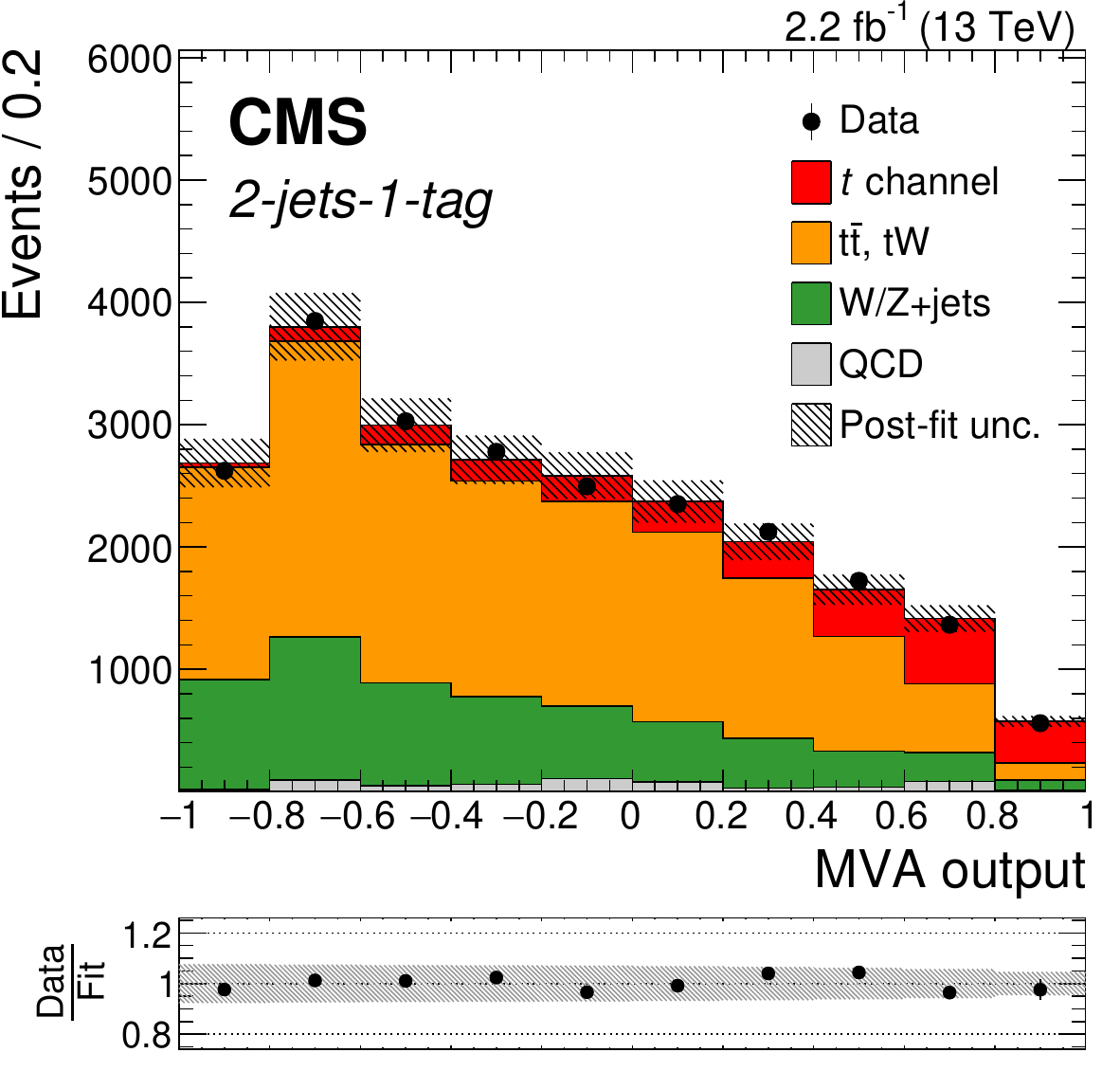}
\includegraphics[height=1.8in]{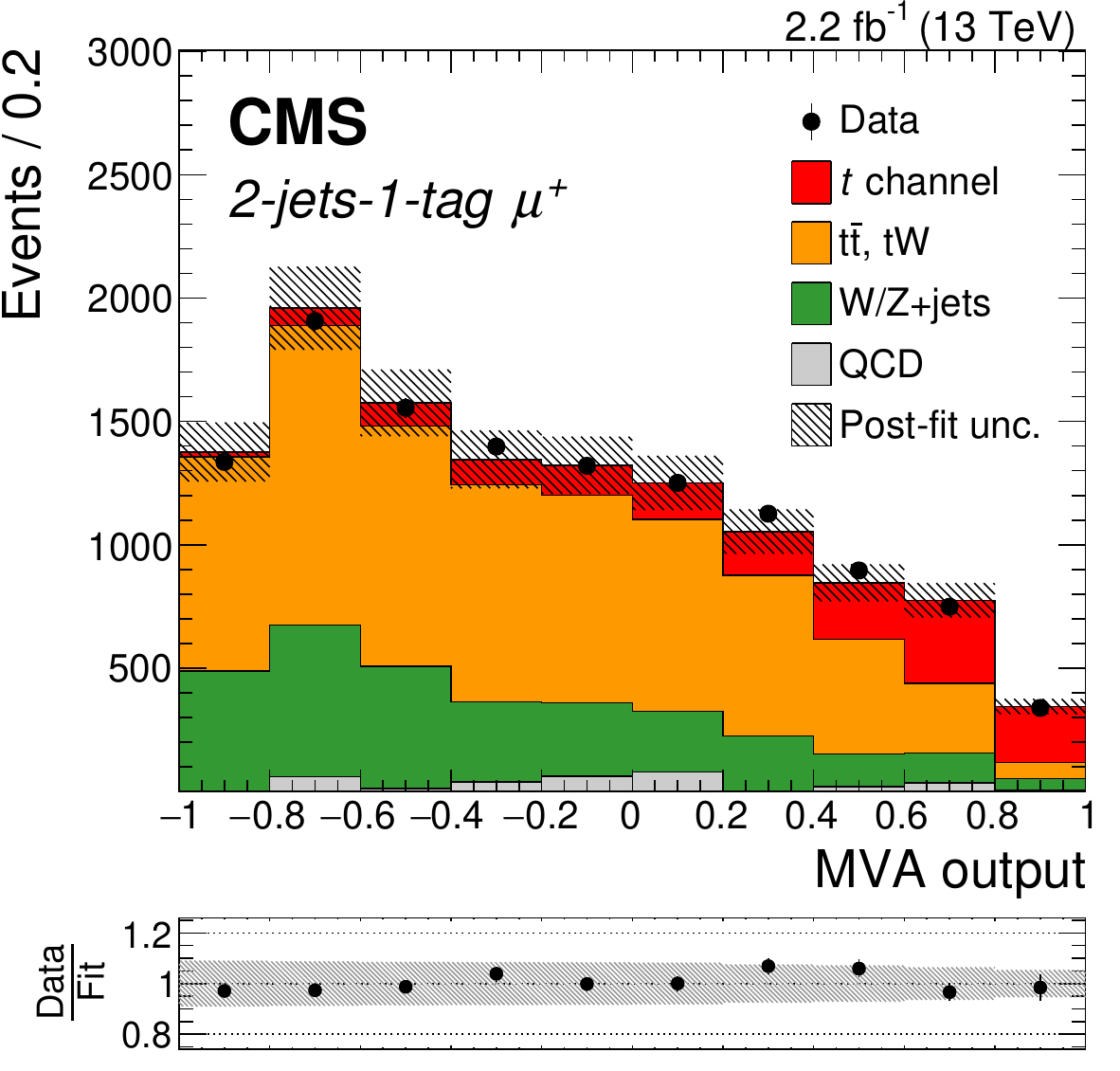}
\includegraphics[height=1.8in]{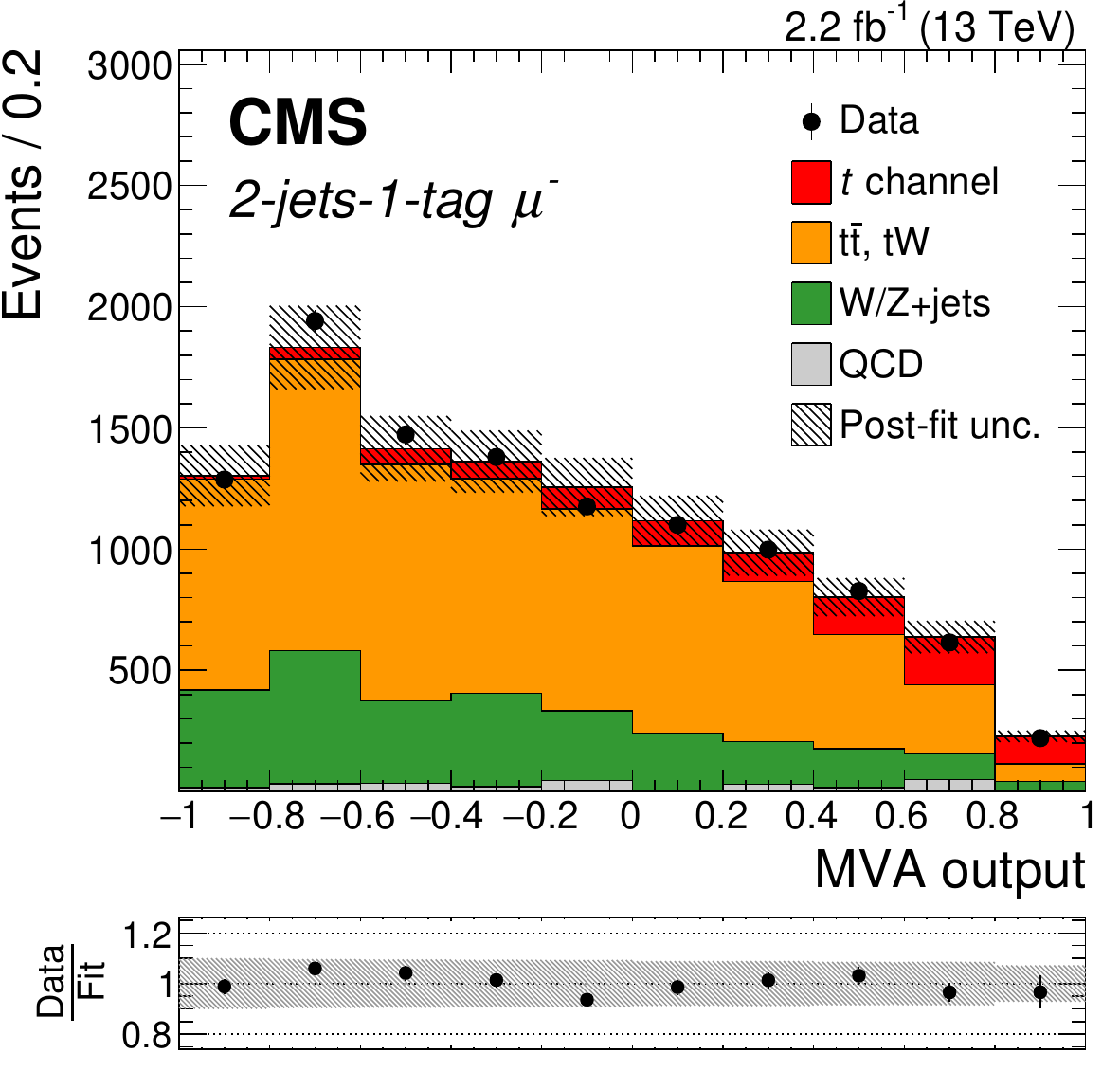}
\caption{Distribution of the multivariate classifier in 2-jets--1-tag signal region. The prediction is scaled to the result of the maximum-likelihood fit. The distribution for the inclusive selection is shown on the left, while the ones for positively and negatively charged muons are shown in the middle and on the right, respectively. Good agreement between simulation and data is observed~\cite{paper}.}
\label{fig:sr}
\end{figure}

\begin{figure}[htb]
\centering
\includegraphics[height=3in]{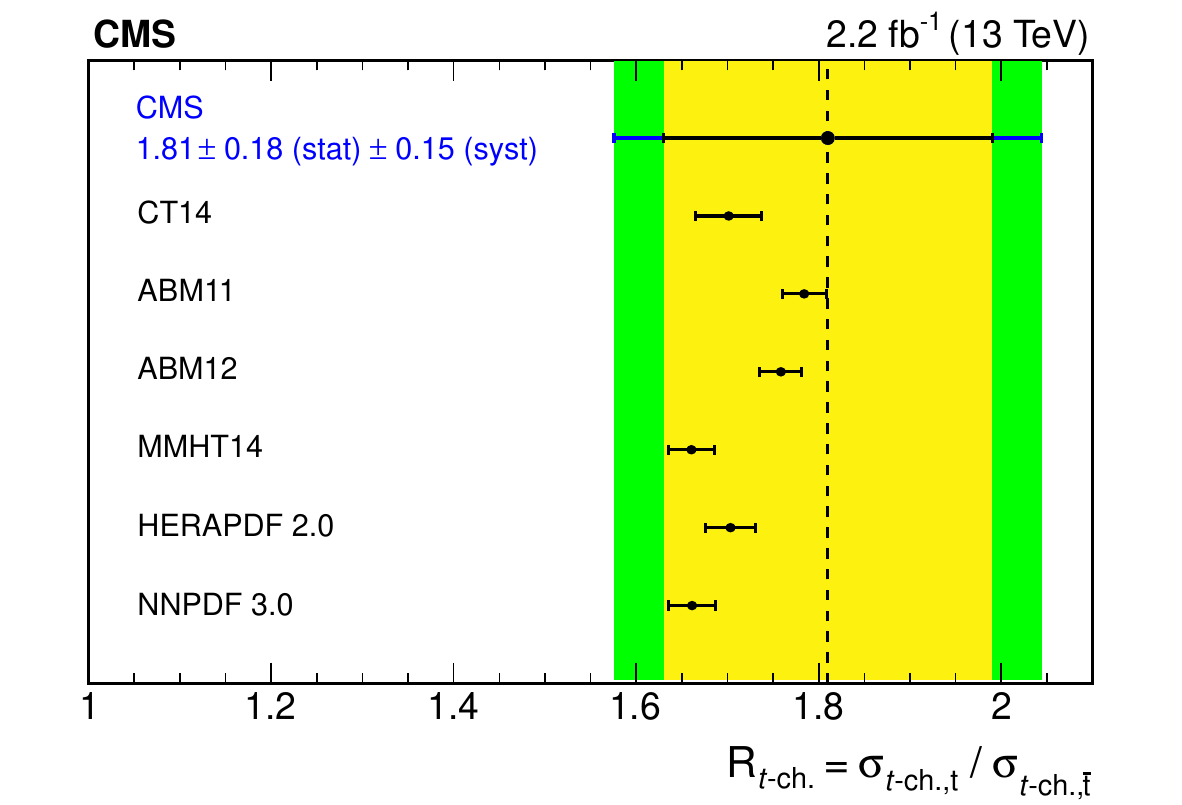}
\caption{Measured ratio of top and antitop quark production and the prediction from different PDF sets. Uncertainties on the predictions account for the choice of factorization and renormalization scales, the top quark mass and statistical uncertainties. The CMS result was measured using the NNPDF 3.0 PDF set~\cite{paper}.}
\label{fig:ratio}
\end{figure}

The analysis uses proton-proton collision data recorded in 2015 at the LHC corresponding to an integrated luminosity of $L\,=\, 2.2\,\textrm{fb}^{-1}$. The collisions are recorded with the CMS detector~\cite{cms}. Because of the low off-diagonal elements of the CKM matrix the top quark decays in nearly all cases into a W boson and a bottom quark. Thus, the decay of the top quark is characterized by the decay of the W boson. In this analysis only leptonic W bosons are studied, where decays into tau leptons are only considered if the tau lepton decays further into a muon. The final state of the process consist of an light-flavored quark, which is usually emitted in forward direction, a bottom quark, a muon and a neutrino. A second bottom quark can also be present (see Fig.~\ref{fig:fdia}), but it often fails the detector acceptance.\\
Events are selected which contain exactly two jets, where only one is identified as originating from a bottom quark, and one isolated muon. As the neutrino escapes the detector without any interaction only the transverse component can be indirectly measured using the $p_{\mathrm{T}}$ balance of the event. The W boson is reconstructed by constraining its mass to the literature value and solving the resulting quadratic equation. Finally, the top quark is reconstructed by adding the four-momenta of the W boson and the jet originating from the bottom quark.
One important background of the analysis is the contribution from QCD multijet events. Since it cannot be reliable estimated from simulation a data-driven approach is used. The distribution of the transverse W boson is used to distinguish QCD and non-QCD processes. A fit to this distribution is performed, where the two free parameters are the normalization of the QCD and non-QCD templates. The result of the fit is shown in Fig.~\ref{fig:qcd}. A cut on $m^{\mathrm{W}}_{\mathrm{T}}\,>\,50\,\mathrm{GeV}$ is applied to further reduce the QCD contribution in the following.
The contribution from all other processes is estimated with Monte Carlo simulations. Expected and observed event yields in the signal region are shown in Table~\ref{tab:yields}.
\begin{table}[t]
\begin{center}
\begin{tabular}{ c|c|c } 
  Process & $\mu^{+}$& $\mu^{-}$ \\
  \hline
  Top quark ($\mathrm{t}\bar{\mathrm{t}}$~and tW) & $6837 \pm13$ & $6844 \pm 13$ \\
  W+jets and Z+jets & $2752 \pm 82 $ & $ 2487 \pm 76 $ \\
  QCD multijet &$ 308 \pm154$ & $ 266 \pm 133$ \\
  \hline
  Single top quark $t$-channel&$1493 \pm 13 $ & $ 948 \pm 10 $ \\
  \hline
  Total expected& $11390 \pm 175$ &$ 10545 \pm 154 $\\
  \hline
  Data&11877&  11017  \\
\end{tabular} 
\caption{Event yields in the signal region for positively and negatively charged muons. The expected numbers are taken from Monte Carlo simulations, where the uncertainty corresponds to the size of the simulation samples. Only the QCD multijet contribution is estimated using data and an uncertainty of $50\%$ is used.}
\label{tab:yields}
\end{center}
\end{table}\\
A neural network is trained in the signal region to combine the separation power of several variables into one discriminating variable. The most important input variables are the pseudorapidity of the light-flavored quark and the reconstructed top quark mass. A maximum-likelihood fit is performed to the output distribution of the neural network. The signal is unconstrained and the background processes have a specific prior with a width according to the uncertainty of their theoretical cross section. The fit is performed simultaneously in the signal region and two control regions to constrain the top quark background. The resulting distribution in the signal region is shown in Fig.~\ref{fig:sr}.
The measured cross sections are:
\begin{eqnarray*}
\sigma_{t\textrm{-ch.,t}} & = & 154 \pm 8\,\rm{(stat)} \pm 9\,\rm{(exp)} \pm 19\,\rm{(theo)}  \pm 4\,\rm{(lumi)}\,\rm{pb}\\
\sigma_{t\textrm{-ch.,}\bar{\textrm{t}}} & = & 85 \pm 10\,\rm{(stat)} \pm 4\,\rm{(exp)} \pm 11\,\rm{(theo)}  \pm 2\,\rm{(lumi)}\,\rm{pb} \\
\sigma_{t\textrm{-ch.},\textrm{t}+\bar{\textrm{t}}} & = & 238 \pm 13\,\rm{(stat)} \pm 12\,\rm{(exp)} \pm 26\,\rm{(theo)} \pm 5\,\rm{(lumi)}\,\rm{pb} 
\end{eqnarray*}
It is possible to probe the quark PDF of the initial state as the ratio of top and antitop quark production is sensitive to the ratio of up- and down-type quarks in the initial state. This ratio is shown in Fig.~\ref{fig:ratio} along with the predictions from different PDF sets. All measured observables are in agreement with standard model predictions. Additional details can be found in Ref.~\cite{paper}.

\end{document}